\documentclass[aps,prl,nofootinbib,amsmath,twocolumn,preprintnumbers]{revtex4-1}
\usepackage{amssymb,esvect,amsmath,graphicx,latexsym,amsthm,slashed,eso-pic,hyperref}
\usepackage[hang,flushmargin]{footmisc}

\newcommand{\beq}{\begin{equation}} \newcommand{\eeq}{\end{equation}}
\newcommand{\bea}{\begin{eqnarray}} \newcommand{\eea}{\end{eqnarray}}

\newcommand{\lhsc}{\lambda_{hs}^c}
\newcommand{\lhs}{\lambda_{hs}}
\newcommand{\ns}{N_s}

\begin{document}
\preprint{YITP-SB-18-19}

\title{Unrestored Electroweak Symmetry}

\author{Patrick Meade}
\affiliation{C. N. Yang Institute for Theoretical Physics, Stony Brook University, 
 Stony Brook, NY 11794}

\author{Harikrishnan Ramani}
\affiliation{Berkeley Center for Theoretical Physics, Department of Physics, University of California, Berkeley, CA 94720}
\affiliation{Theoretical Physics Group, Lawrence Berkeley National Laboratory, Berkeley, CA 94720}

\begin{abstract}
The commonly assumed cosmological history of our universe is that at early-times and high-temperatures the universe went through an ElectroWeak Phase Transition (EWPT).   Assuming an EWPT, and depending on its strength, there are many implications for baryogenesis, gravitational waves, and the evolution of the universe in general.  However, it is not true that all spontaneously broken symmetries  at zero-temperature are restored at high-temperature.   In particular the idea of ``inverse symmetry breaking" has long been established in scalar theories with evidence from both perturbative and lattice calculations. In this letter we demonstrate that with a simple extension of the SM it is possible that the ElectroWeak (EW) symmetry was always broken or only temporarily passed through a symmetry restored phase.  These novel phase histories have many cosmological and collider implications that we discuss.  The model presented here serves as a useful benchmark comparison for future attempts to discern the phase of our universe at $T\gtrsim$ a few GeV.

\end{abstract}
\maketitle

\section{Introduction}

Since the discovery of a Standard Model(SM)-like Higgs at the LHC~\cite{Aad:2012tfa,Chatrchyan:2012xdj},  there have been numerous attempts to understand its implications for physics beyond the SM (BSM) and the cosmological history of the universe.  In particular, the nature of the EWPT has been investigated in detail due to its possible connection to EW Baryogenesis~\cite{Kuzmin:1985mm}.  However, the EWPT is an interesting question to study in its own right.   In the SM, using even the simplest original techniques~\cite{Kirzhnits:1976ts,Weinberg:1974hy}, there is an EWPT to an unbroken phase at high temperatures.  Nevertheless, the potential for the Higgs has yet to have been measured precisely enough to determine whether or not there are differences from the SM prediction of a second-order phase transition.   This has driven much work over the past decades, and experimentally determining the shape of the Higgs potential is a compelling driver for future experimental physics programs~\cite{Arkani-Hamed:2015vfh,Contino:2016spe}.   In particular, testing at future colliders whether the phase transition is consistent with being first-order is a way to probe whether EW barogenesis is even possible~\cite{Curtin:2014jma}.  Additionally, if there was a first order EWPT it would necessarily imply new particles that couple to the Higgs which could be discovered by future colliders.  There are also cosmic connections for a first-order EWPT, as it would create a potentially measurable gravitational wave (GW) signal(see e.g~\cite{Grojean:2006bp} and references therein).  This is especially interesting given that we are now in the era of GW astronomy and the frequency range of interest overlaps the sensitivity of next generation GW experiments.

While studying the order of the EWPT provides a compelling research program for high energy experimental physics and GW astronomy, there is an even more basic question that can be investigated.  Did an EW phase transition ever occur in the early universe?  The original techniques for studying finite temperature QFT~\cite{Kirzhnits:1976ts,Weinberg:1974hy} gave a robust mechanism for restoring symmetries at high-T, to the point that it became almost common lore.   This stemmed from the fact that scalars acquire a thermal mass at leading order of the form
\begin{equation}\label{eqn:highTrestore}
V(\phi,T)\supset \eta \phi^2 T^2,
\end{equation}
where $\eta$ represents the coupling of the scalar $\phi$ to other particles or itself.  Eventually at high-T this term would dominate any negative quadratic combination responsible for spontaneous symmetry breaking and restore the resulting symmetry.  However, it was also pointed out in~\cite{Weinberg:1974hy} that symmetries need not be restored at high-T, and there are other symmetry restoration/non-restoration patterns that could be realized in nature, i.e. Inverse Symmetry Breaking (ISB) or Symmetry Non-Restoration (SNR).  While at first these were phenomena were discovered perturbatively, they have since been realized on the lattice~\cite{Jansen:1998rj,Bimonte:1999tw} and there is sufficient additional evidence that this possibility is now on firm theoretical footing.   Additionally as reviewed in~\cite{Weinberg:1974hy} systems exist in nature that demonstrate this phenomenon, such as Rochelle Salts and particular Liquid Crystal systems~\cite{Bajc:1999cn}.   ISB and SNR have also been postulated as solutions to various problems, for example the matter-antimatter asymmetry with persistent CP violation~\cite{Mohapatra:1979qt,Mohapatra:1979vr}, and to avoid Monopoles~\cite{Dvali:1995cj,Salomonson:1984rh,Langacker:1980kd} and Domain Walls~\cite{Dvali:1995cc} in GUTs.

In this letter we will show that with a simple scalar extension of the SM, a non-restoration phase can occur for the EW symmetry.   The existence of such a phase has a number of implications experimentally and cosmologically.  First, the cosmological history is very different than the SM.  While it is commonly assumed that all SM particles are massless before the standard EWPT, this is not the case in the phase we describe, and masses {\em increase} with temperature.  This influences how particles decouple and can provide alternative cosmic histories for relic abundances as well as novel equations of state.   Correlated to the temperature dependent VEV, EW sphalerons are inefficient because of the persistent EWSB and models of baryogenesis which utilize them are not viable.  This implies that old mechanism such as GUT baryogenesis are potentially viable and models of EW baryogenesis are not viable.  There also will not be a gravitational wave signal, because there is no first order phase transition.   Finally, because the singlets must couple to the Higgs there can be correlated collider signals.  However, as we demonstrate, these collider signals can be  invisible to future colliders potentially invalidating no-lose theorems for testing the EWPT~\cite{Curtin:2014jma}.

\section{Model}

To demonstrate SNR for EWSB we exploit the same term used for symmetry restoration~(\ref{eqn:highTrestore}).  As pointed out in~\cite{Weinberg:1974hy}, mixed quartics can be negative and this term can also cause SNR as well as restoration.  In this letter we consider a SM singlet $s$ transforming in a vector representation of an $O(N_s)$ global symmetry coupling to our Higgs through the following Lagrangian
\begin{equation}
\mathcal{L}=\mathcal{L_{\text{SM}}}-\frac{\mu_s^2}{2} s^2-\frac{1}{4} \lambda_s s^4 -\frac{\lambda_{h s}}{2} h^2 s^2.
\end{equation}

The quartics $\lambda_s$ and the SM Higgs quartic $\lambda$ must be positive, but the mixed quartic $\lambda_{hs}$ can be negative.  However, to avoid a negative runaway direction, $\lambda_{hs}$, is is bounded such that
\begin{equation}
\lambda_{h s} \ge -\sqrt{ \lambda_s \lambda}.
\label{eqn:lambdahsbound}
\end{equation}

\begin{figure}[htbp] 
   \centering
   \includegraphics[width=3in]{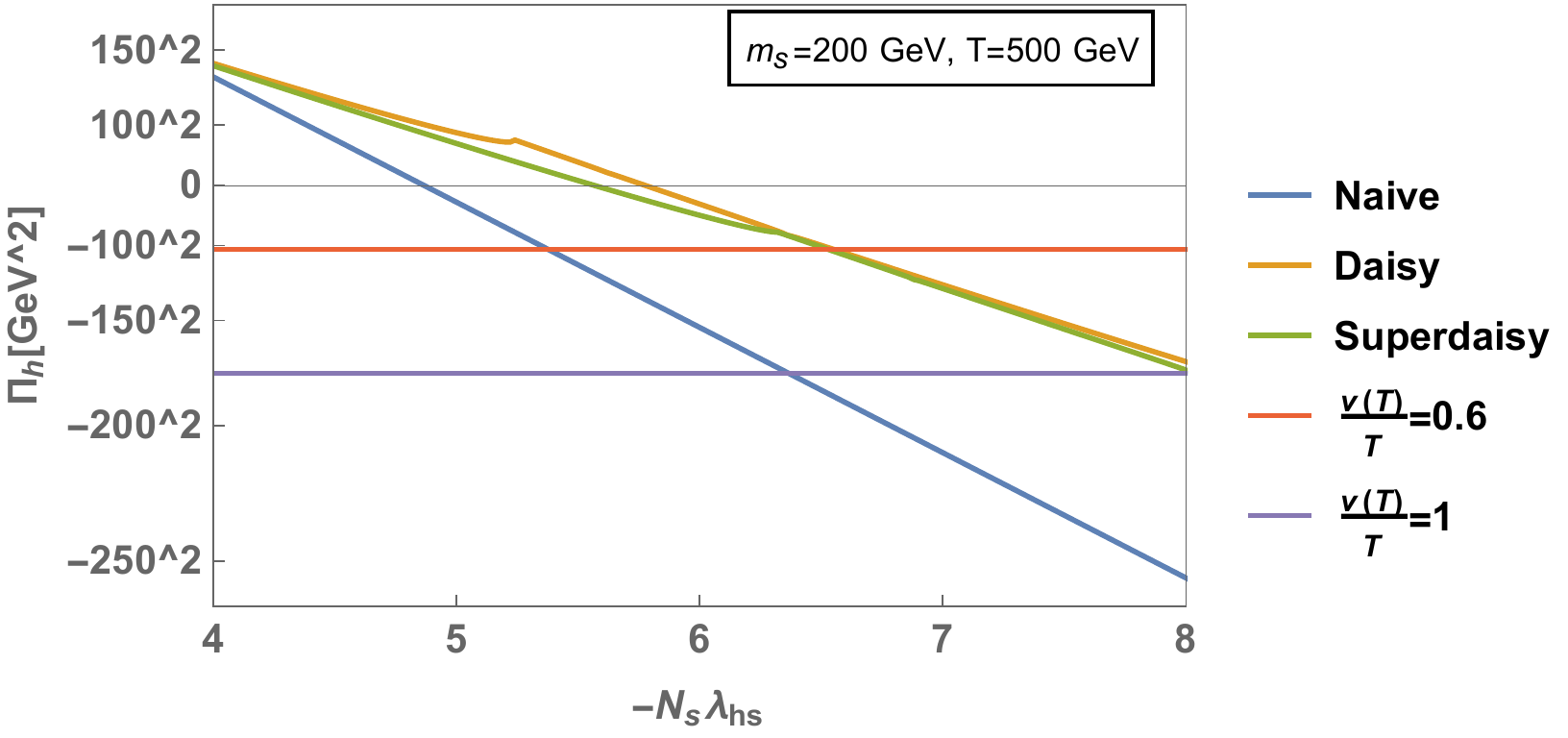} 
   \caption{For an example point with $m_s=200\,\mathrm{GeV}$ and $N_s=600$ we show comparisons between the naive expectations from Eqn.~\ref{eqn:pih} and the  Daisy and SuperDaisy contributions, for $T=500\,\mathrm{GeV}$. 
   }
   \label{fig:piresum}
\end{figure}

\begin{figure}[htbp] 
   \centering
   \includegraphics[width=3in]{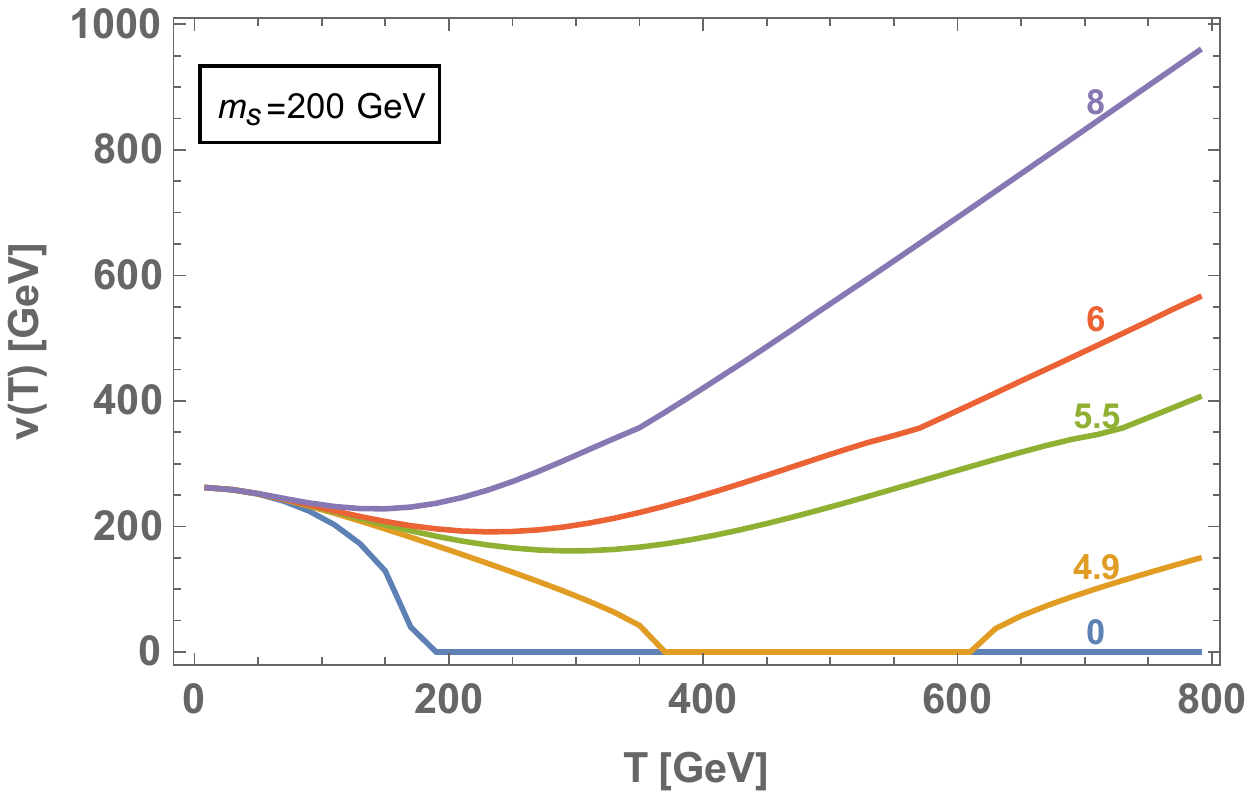}    
   \includegraphics[width=3in]{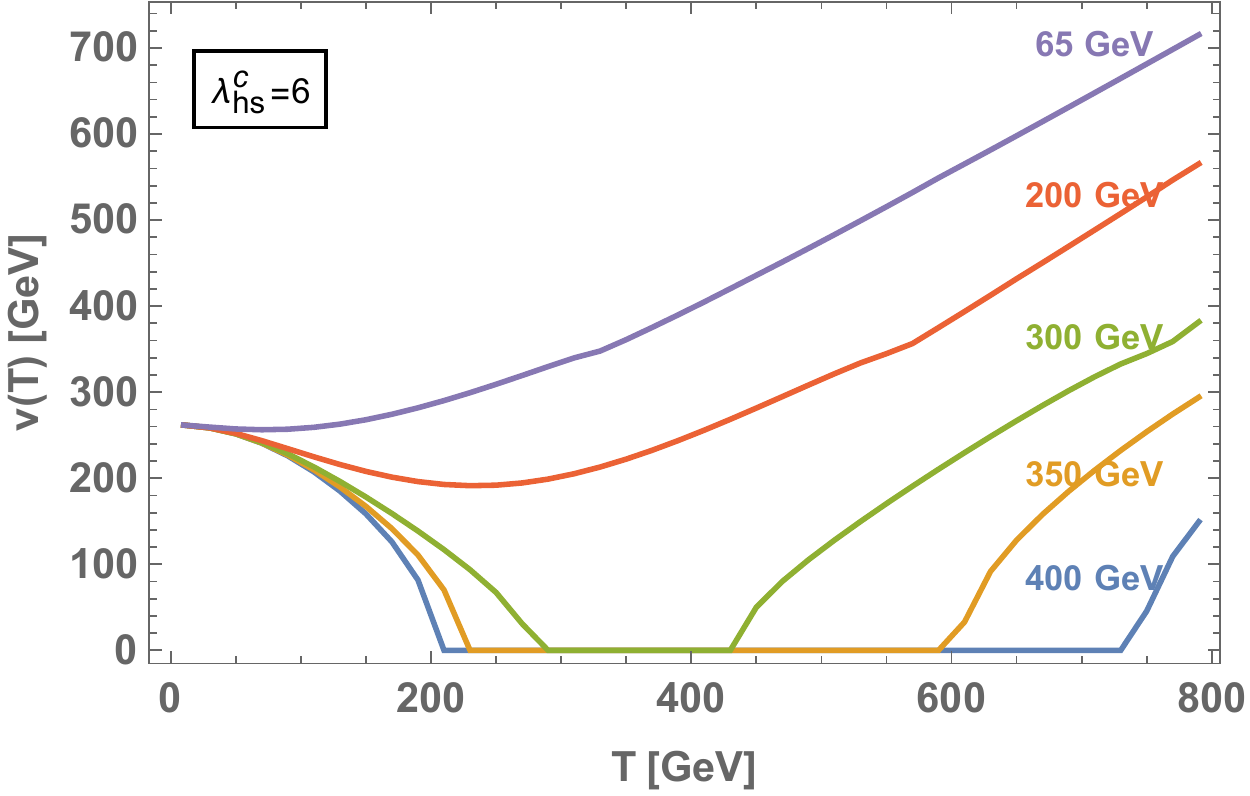} 
   \caption{Top: The temperature dependent VEV, $v(T)$ for different values of $\lambda_{hs}^c$ demonstrating different phase histories as a function of the temperature, $T$.  Bottom: The temperature dependent VEV for a fixed $\lambda_{hs}^c$ and different values of $m_s$. $N_s=600$ to exhibit the large $N_s$ limit.}
   \label{fig:phases}
\end{figure}

Therefore a contribution to~(\ref{eqn:highTrestore}) for the Higgs can be negative from a negative $\lambda_{hs}$ in this range and possibly allow for SNR.  Nevertheless, to achieve SNR the negative contribution must outweigh the usual positive contributions.  For the Higgs there are positive contributions to its thermal mass from SM particles and the Higgs itself.   At leading order in the high temperature limit the thermal masses of $h$ and $s$ are given by,
 \begin{eqnarray} 
 \label{eqn:pih}
\Pi_h &=& T^2\left(\frac{\lambda_t^2}{4} +\frac{3g^2}{16} +\frac{g'^2}{16} +\frac{\lambda}{2}+N_s\frac{\lambda_{hs}}{12}\right)\\
\label{eqn:pis}
\Pi_s &= &T^2\left((N_s+2) \frac{\lambda_s}{12} + \frac{\lambda_{hs}}{3}\right). 
\end{eqnarray}
The simplest scalar extension, where $N_s=1$, requires $\lambda_{hs} \ll -1$ for SNR to overcome the top Yukawa contribution in~(\ref{eqn:pih}).  This requires that $\lambda_{s}$ must be non-perturbative to satisfy~(\ref{eqn:lambdahsbound}), which is why this region is usually excluded from interest~\cite{Curtin:2014jma}.   However, by taking $N_s \gg 1$,  it is simple to simultaneously achieve SNR through~(\ref{eqn:pih}) while maintaining perturbativity of the couplings.  Additionally in this scaling limit, at high-T the theory remains only in a partially ordered phase where $\langle s \rangle = 0$.  This is because the large $N_s$ limit ensures symmetry restoration at high-T from ~(\ref{eqn:pis}) regardless of the sign of $\mu_s$.  The existence of this phase of the SM has been shown thus far using the leading order expansion in high-T, and large $N_s$.  In this regard, the phase is robust as the couplings can be shown to be as small as required for perturbative control, similar to the arguments made in large-$N$ for the $O(N_1)\times O(N_2)$ model in~\cite{Orloff:1996yn}.  However, this can be extended to finite $N_s$, beyond the Daisy limit, and including finite mass effects using the methods developed in~\cite{Curtin:2016urg} without changing the conclusions.  Additionally it is interesting to note that some of the typical problematic finite temperature infrared effects for gauge theories that undergo spontaneous symmetry breaking are different in this phase since there is never a symmetry restored limit.  In Figure~\ref{fig:piresum} we demonstrate how the effective $\Pi_h$ can be quite different than the naive expectation of Eqn.~\ref{eqn:pih}.
In particular, by including these higher order effects there are many possible thermal histories of our universe that can be realized.   For instance, either $m_s$ effects or varying $N_s \lhs$ can result in the EW symmetry having SNR or other phase histories where one passes from an ordered to disordered and then back to an ordered phase as shown in Figure~\ref{fig:phases}, which we refer to as a temporary restoration (TR) phase.  This can be understood through either the decoupling of thermal effects, or how large of $\Pi_h$ results from the coupling constant.  It should be emphasized that TR could extend for a very long period if $m_s$ is very large.  Additionally, more complicated phase histories could exist alternating between SR and NR if additional scalars are appropriately added to the model.

\section{Phase Stability}

The existence of SNR, or more complicated phase histories, for the simple model discussed here is robust when looked at from many different vantage points including: RGE stability, thermal decoupling and thermal fluctuations. In the large-$N_s$ limit we can define an effective 't Hooft coupling for the theory
\begin{equation}
\lambda_{hs}^c\equiv \vert \lambda_{hs}\vert N_s
\end{equation}
which helps organize perturbation theory more clearly.  In particular, our condition for SNR will just be that $\lhsc$ is greater than some fixed $\mathcal{O}(1)$ value depending on the parameter point.  With this coupling definition, our condition for stability of the potential at zero-temperature is given by
\begin{equation}
\lambda_s\geq \left(\frac{\lhsc}{\ns}\right)^2 \frac{1}{\lambda},
\end{equation}
which clearly allows for pertubatively small $\lambda_s$.  Under RGE evolution for large $N_s$ the 1-loop $\beta$-functions reduce to
\begin{equation}\label{bfunctions}
\begin{split}
\beta_{\lambda} = \beta_{\lambda}^{SM} \quad \beta_{\lambda^c_s} = \frac{1}{16\pi^2} 2( \lambda^c_{hs})^2
\\ \beta_{\lambda^c_{hs}} = \frac{\lambda^c_{hs}}{16 \pi^2}(12\lambda+6 y_t^2-\frac{3}{2}g_1^2-\frac{9}{2} g_2^2),
\end{split}
\end{equation}
where we have defined $\lambda_s^c\equiv \lambda_s \ns^2$.   From this basic 1-loop structure it can clearly be seen that the phase structure is robust against RGE evolution.  We have numerically verified using 3-loop RGEs that the couplings can remain perturbative to the Planck scale in this model.    However, it is important to note that there is still interesting phenomenology that can come out of the solution to the RGE evolution.  The effective finite temperature VEV scales as
\begin{equation}
v(T)\sim \sqrt{\frac{\mu^2-\Pi_h(T)}{\lambda(T)}}
\end{equation}
from which we can define the ratio 
\begin{equation}
\kappa(T)\equiv v(T)/T
\end{equation}
which strongly depends on the value of $\lambda(T)$.  In this notation we are defining the thermal $\lambda(T)$ as containing the zero temperature running coupling constant, as well as the effective change to the quartic from the thermal potential.  It is well known that the temperature dependent contributions to $\lambda$ are small which naively limits the size of $\kappa$ since $\Pi(T)\sim T$.  Nevertheless, at high temperatures the appropriate RGE scale should be $\mu_R\sim T$ and the zero-temperature running therefore is important.  In particular at large $N_s$, the $\beta$-function \ref{bfunctions} for $\lambda$ is SM-like and therefore $\lambda(T)\sim 0$ at large $T$ as with the usual metastibility story in the SM.  Therefore, large values of $\kappa$ are achievable which can cause a number of interesting phenomenological consequences that we will discuss later.  However, there are self-consistency limits on the phase which bound  $\kappa$ from above.  Despite the SM RGE argument, $\kappa\sim\mathcal{O}(1)$ can be maintained to arbitrary high scales from the $1/\ns$ effects in the $\beta$-function for $\lambda$ while all other couplings remain under perturbative control as well.   In all plots in this paper we have chosen $\lambda(T)=\lambda(0)$ for simplicity, but a more dedicated study of the parameter space with all RGE effects for a dynamical scale would be an interesting direction to pursue.

If this phase is robust, then we must also check that the fields stay in thermal equilibrium validating the ansatz of equlibrium local thermal field theory used in describing the phase in the previous section.  For our purposes here, it is sufficient that $S$ and $h$ maintain equilibrium since these are the fields which drive the SNR phase.  If SM fields were included this would only serve to increase the allowed parameter space for SNR.  To establish the validity of thermal equilibrium we can compare the various reaction rates $\Gamma_h~:~(h+h \leftrightarrow h+h)$, $\Gamma_{hs}~:~(h+h \leftrightarrow s+s)$, $\Gamma_s~:~(s+s \leftrightarrow s+s)$, and permutations thereof to Hubble scale, $H$.   
In the large $T$ and large $N_s$ limit, the masses of the particles (using only the leading order contributions to the appropriate $\Pi$) scale as
\begin{equation}
m_h\sim \left(\frac{\lhsc}{12} \right)^{1/2}T \; \mathrm{and}\; m_s\sim \mu_s.
\end{equation}
Therefore at large temperatures all scalars can be treated as relativistic.    The various mixed quartics contribute to $\Gamma_h,\Gamma_{hs},$ and $\Gamma_s$, but there are also trilinear interactions from the Higgs VEV that matter given the scaling of $v(T)$ for SNR.  The trilinear couplings in the large $T$ limit scales as
\begin{equation}
g_{hhh}\sim \lambda \kappa T\;\; \mathrm{and} \;\;g_{hss}\sim \frac{\lhsc}{N_s}\kappa T.
\end{equation}
Naively the scalings of the various couplings imply that the $\Gamma_{hs}$ reaction is in most danger of falling out of equilibrium since one can't formally take the infinite $N_s$ limit.   This would be disastrous of course since then the SNR would not exist in the first place. However, this is not surprising since in this limit
\begin{equation}
H\sim g_*^{1/2} \frac{T^2}{M_{pl}}\sim \frac{N_s^{1/2} T^2}{M_{pl}}
\end{equation}
and even a simple scalar $\phi^4$ interaction wouldn't stay in equilibrium with itself in the infinite $N_s$ limit because of the infinite contribution to the Hubble rate.  Therefore the numerics do matter, and qualitatively taking the $N_s\sim \mathcal{O}(100)$ is more than sufficient to maintain perturbativity in the finite temperature QFT calculations {\em and} allow for thermal equilibrium over a wide range of temperatures.  If $N_s$ is lowered even further the range of temperatures where equilibrium holds is only enlarged, and for the $h\rightarrow s$ equilibrium which provides the strongest constraint, the reactions are in equilibrium for 
\begin{equation}
T \lesssim \left(\frac{\lambda^2(\lhsc)^2\kappa^4}{N_s^{5/2}}\right) M_{pl}.
\end{equation}
This can easily be for $T<T_{GUT}$ depending on the parameter choices, which is the standard decoupling temperature limit for weakly interacting relativistic particles.  Once the temperature of the universe drops to the point where $h$ and $s$ are non-relativistic then they decouple in the usual fashion.  Therefore for $T$ between $\mathrm{max}(m_s,m_h)\lesssim T\lesssim T_{GUT}$ thermal equilibrium can be maintained, with the upper limit being reduced for extremely large $N_s$.  The lower limit unfortunately cannot be reduced further given the parameters of the Higgs sector at zero temperature.

Another concern is the possibility that even though the VEV of the the Higgs scales as the temperature, $T$, the inherent scale of thermal fluctuations is also of order $T$, so do thermal fluctuations take us out of this new SNR vacuum?  Normally this question is asked in the context of a phase transition~\cite{Kibble:1976sj}, where the correlation length $\xi$ is used to compared the difference in free energy density, $\Delta f$ between the broken and unbroken vacuums.  If the free energy $\Delta F \gg T$ then fluctuations back to the unbroken vacuum become highly improbable.  In our case
\begin{equation}
\Delta F\sim \xi^3 \Delta f \sim \frac{(\lhsc)^{1/2}}{\lambda} T\gg T
\end{equation}
so the SNR vacuum is the preferred vacuum that the Higgs stays in.

\section{Cosmology}
There are a number of cosmological differences for SNR or TR phases compared to the usual symmetry restored (SR) phase.  Some of these effects are due to the VEV of the Higgs not vanishing, while others are more connected to the $\kappa$ parameter space. \\
\textbf{Gravitational Waves:}
One simple cosmological consequence of SNR is that in the absence of a phase transition there will be no gravitational wave signal.  However, there will be a difference in how this arises compared to the usual SM statement of a 2nd order phase transition.  In our model, there {\em can} be shifts to the Higgs couplings relative to the SM and measurable at colliders (discussed in the next section) that could naively imply a gravitational wave signal and therefore future GW observatories would be useful to distinguish the phase of the early universe~\cite{Caprini:2015zlo}.  In the more complicated TR history there are two phase transitions, however, there are no gravitational waves because the phase transitions are second order.\\
\textbf{Electroweak Sphalerons and Baryogenesis:}
EW sphalerons are often a key ingredient in models of Baryogenesis ranging from models of EW Baryogenesis to Leptogenesis.  This is due to the fact that they provide a $B+L$ violating process, and are useful as a SM source of baryon number violation or to reprocess a lepton asymmetry into a baryon asymmetry.  However, they critically rely on the fact that the EWS is restored at higher temperatures since they are exponentially suppressed by $\sim e^{-4\pi/\alpha_w}$ at zero-temperature.  However, in a model with SNR or TR, the effective suppression is modified compare to the usual symmetry restoration because
\begin{equation}
\Gamma_{sph}\sim d_1 (\alpha_w M_W(T))^4 \exp\left(-\frac{d_2 \kappa}{\alpha_{w}}\right)
\end{equation}
with constants $d_1\lesssim \mathcal{O}(1)$ and $d_2 \gtrsim \mathcal{O}(1)$, implying that in SNR or TR with $\kappa\gtrsim \mathcal{O}(1)$, the exponential suppression persists at high $T$.  Only in models of TR would the sphalerons be temporarily active, and then it is a question of model parameters as to whether or not there is a sufficient time to generate and/or process a successful baryon asymmetry.  There are of course a variety of models that don't require/are unaffected by EW sphalerons, e.g. Affleck-Dine baryogenesis~\cite{Affleck:1984fy} for a field with net  $B-L$ and an appropriate decay.  On the other hand, EW sphalerons strongly constrain models which don't generate a net $B-L$, e.g. $SU(5)$ GUTs, since they wash out $B+L$ generated by out of equilibrium decay of GUT particles. SNR or TR would allow for such a model and for direct $B$ violation at high scales to be a possible Baryogenesis explanation.  In particular this could allow for compatibility with existing Nucleon Decay and Oscillation experiments which is an interesting direction to pursue theoretically and experimentally.  Additionally, since sphalerons in a SR phase are out of equilibrium for $T>10^{12}\,\mathrm{GeV}$, it isn't necessary for SNR or TR phenomena to persist to the GUT scale.

\textbf{Thermal Evolution:}
For the SNR/TR phases discussed here there can be changes to the overall thermal evolution of the universe through contributions to $H$, novel equations of state for particles who obtain mass through the Higgs, and changes in decoupling/recoupling to the thermal bath.  For instance, the addtional large $N_s$ additional scalars act as radiation and contribute $\rho_S \sim N_s T^4$ to the energy density at early times in addition to the SM radiation bath, and can overwhelm the SM contribution depending on $N_s$.  However, since temperature dependent effects don't extend below $T_{EW}$ it isn't necessary for $m_s$ to be light in which case the effects can completely disappear well before any measures such as $N_{eff}$ are relevant.   There can also be a contribution to the DM relic density, $\Omega_{DM}$,  from $S$, but these contributions to $\Omega_{DM}$ can be tuned away and other particles can serve as the DM in simple extensions that don't alter the SNR/TR phases.  One potentially interesting contribution to $H$ from a SNR/TR is via the putative vacuum energy contribution from the Higgs potential which is given by $\rho_V = -\frac{1}{48} \lambda_{hs}^c \kappa^2 T^4 $.  This only potentially matters for very large $\kappa$ and as such acts as a self-consistency constraint on the large-$\kappa$ limit. It is also a potentially novel early universe quintessence-like setup given the $T$ dependence.  There are no constraints on $\dot{w}_V$ at high $T$ and since our contribution doesn't persist to low $T$ it is not constrained.  There could also be additional unknown contributions to the vacuum energy which could modify any conclusion reached about $\rho_V$ alone.

Another interesting cosmological effect concerns the equation of state for any particle which obtains its mass from the Higgs.  For a particle, $i$, with a zero-temperature mass contribution given by $m_i\sim c_i v$, where $c_i$ is the coupling to the Higgs field, at finite temperatures $m_i(T)\sim c_i \kappa T$.  If $\kappa$ is very large the particle is non-relativistic and doesn't contribute to $g_{*}$, so for instance the massive EW gauge bosons and top quarks drop out of the SM plasma at early times and alter the evolution of $g_*^{SM}$.  There can also be non-standard evolutions of $w$ as the universe cools.  In typical cosmologies particles either stay relativistic, or change from relativistic to non-relativistic matter. However, for large $c_i \kappa$, the equation of state can change from NR matter at high $T$ to quasi radiation near $T_{EW}$ to NR matter at low scales.  For moderate $c_i \kappa$, $w$ can asymptote at high $T$ to a state in between radiation and matter before ultimately acting as NR matter at low temperatures.  Depending on the model there could be interesting potential early matter dominated scenarios for a DM particle that would alter structure formation, or provide a different scaling of $H(a)$.

 The symmetry restored thermal history is known to keep all particles in thermal equilibrium from around the electroweak scale to $10^{15}$ GeV. All the particles in the SM except for the neutrinos are in thermal equilibrium through contact interactions, as long as the masses are $\mathcal{O}(T)$ or lower, the reaction rates are identical to the symmetry restored vanilla cosmology. The Hubble rate on the other hand, could be different due to the extra degrees of freedom that might be present. As long as $g_*\sim N_s$ is not astronomically larger than $g_*^{SM}$ (as discussed in the previous section, extremely large $N_s$ result in decoupling) we should expect very similar cosmology for $\kappa \sim 1$ . However large $\kappa$ could produce novel effects.
 
Thermal effects can also potentially modify freeze-in and freeze-out calculations which have potential effects on the abundance of both SM and DM particles.  Neutrinos are a familiar example via their decoupling caused by  the massive $W$ and $Z$ bosons in the SM.  In the large $\kappa$ limit, neutrinos are in equilibrium for
\begin{equation}
T \le 10^{-2} \frac{M_{\text{pl}}}{g_*^{1/2} \kappa^4}. 
\end{equation}
unlike in the SM where they were in equilibrium for any $T$ below $\sim 10^{16}$ GeV.  For very large $\kappa$ this could lead to neutrinos recoupling at a relatively low temperature (still well above the EW scale) and then decoupling again at around an MeV.  Unfortunately this won't be measurable in $N_{eff}$ since even in the extreme $\kappa$ limit the neutrinos rapidly thermalize around $T_{EW}$ where the SNR/TR effects disappear, but it still provides an alternate thermal history for neutrinos.  Novel DM scenarios are also possible.  DM freeze-out calculations typically have small effects given the scales involved, but this could be altered in a model where the DM mass has a SNR contribution from a sector which is not the Higgs.  For freeze-in DM scenarios, bath particles still in equilibrium can be Boltzmann suppressed at temperatures well above their zero temperature masses, and this could result in a lower freeze-in yield for feebly interacting massive particles (FIMPs), which needs to be compensated by larger couplings for FIMPs to the standard model.

\section{Colliders}
The collider phenomenology of this model is very similar to other models with a singlet and a $Z_2$ symmetry, such as the nightmare scenario in~\cite{Curtin:2014jma}, as its collider phenomenology is governed by the singlet-Higgs coupling.  For $m_s \ge \frac{m_h}{2}$ the interesting collider phenomenology arises in shifts to the Z-Higgs coupling, $\delta_{Zh}$, and in the triple Higgs coupling, $\delta_{h^3}$. For lighter $m_s$, there is also the possibility of direct production through an off-shell Higgs, $\sigma_{h* \rightarrow SS}$.  However, the additional multiplicity of scalars, $N_s$,  gives an additional scaling in the collider observables compared to~\cite{Curtin:2014jma}:
\begin{equation}
 \quad  \delta_{Zh} \sim N_s \lambda_{hs}^2 \quad \sigma_{h* \rightarrow SS} \sim N_s \lambda_{hs}^2 \quad \delta_{h^3} \sim N_s \lambda_{hs}^3.
 \end{equation}
When recast in terms of $\lhsc$ the observables scale as
\begin{equation}
 \quad  \delta_{Zh} \sim \frac{(\lambda^c_{hs})^2}{N_s} \;\sigma_{h* \rightarrow SS} \sim  \frac{(\lambda^c_{hs})^2}{N_s} \;\delta_{h^3}\sim \frac{- (\lambda^c_{hs})^3}{N_s^2} 
 \end{equation}
which shows that in the large $N_s$ limit all the observables  return to the SM value.  The regime $m_S \le \frac{m_h}{2}$ is also potentially viable in the large $N_s$ limit as the contribution to the Higgs partial width
\begin{equation}
\Gamma_{h\rightarrow ss} \sim \frac{(\lambda^c_{hs})^2}{N_s}.
\end{equation}
scales away.  However, it is not particularly interesting from the perspective of SNR or TR since the Higgs is already effectively decoupled from the thermal bath for $m_S$ in this range, and thus there are no interesting thermal effects. 

\begin{figure}[htbp] 
   \centering
   \includegraphics[width=3in]{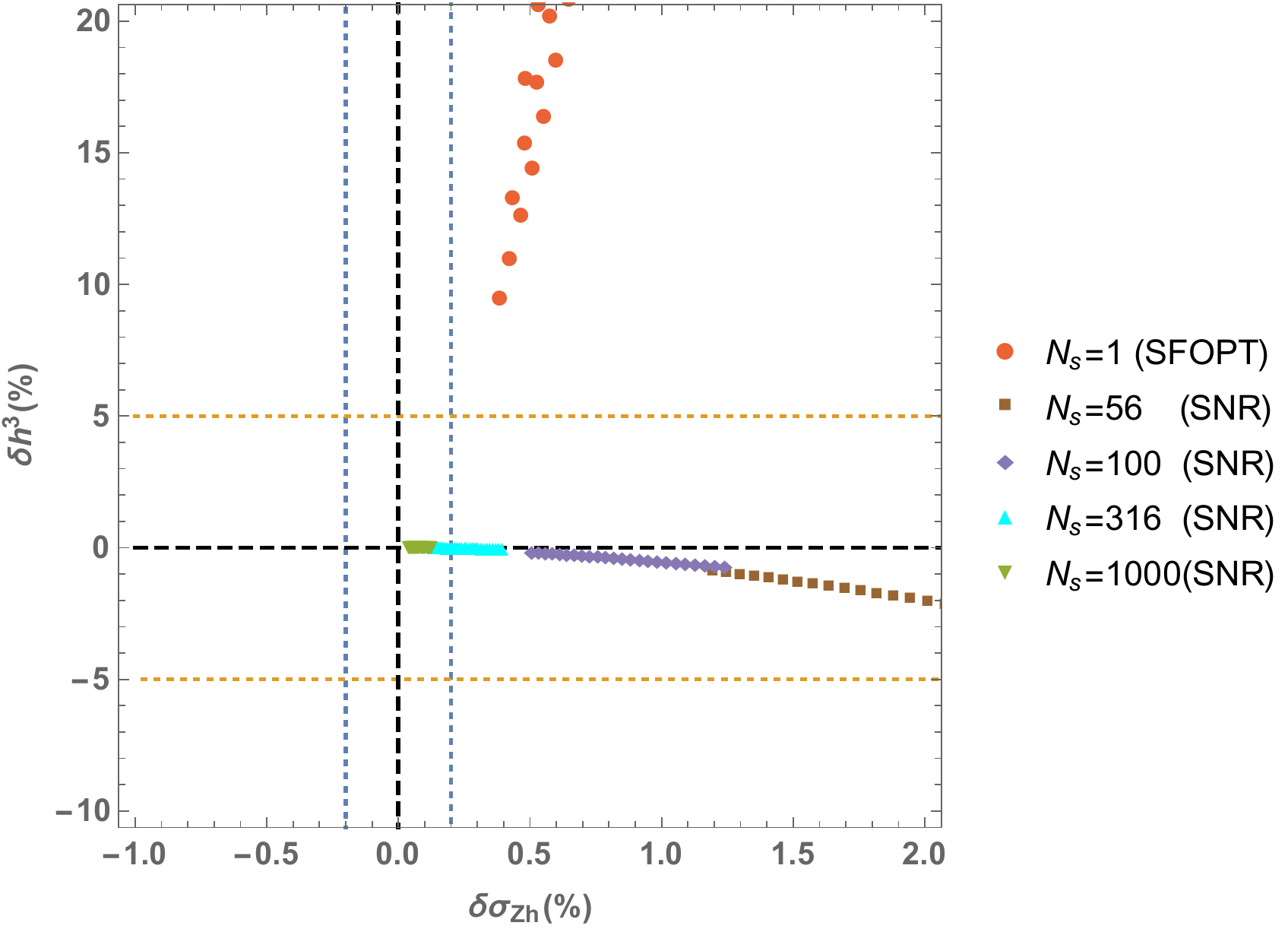} 
   \caption{Comparison between different SNR points and a SFOPT in the space of $ \delta_{Zh}$ and $\delta_{h^3}$, for the onset of SNR and $v_c/T_c=0.6$.   The dashed lines represent projected sensitivities that can be found in~\cite{Curtin:2014jma,Contino:2016spe}; $\delta \sigma_{Zh}=0.2 \%$ and $\delta h^3 =5\%$ }
   \label{fig:colliderplot}
\end{figure}

Given that all the observables scale away in the large $N_s$ limit, it is not obvious what we can learn from colliders about SNR and TR.  In principle, colliders can offer no insight into whether we live in a universe with SNR since it's possible that there is no observable difference with respect to SM Higgs physics.  However, there are interesting questions that still can be investigated when we use SNR or TR as a comparison test with other potential cosmologies.    For instance, for finite $N_s$ is it possible to disentangle SNR from a strongly first order phase transition (SFOPT). In Figure \ref{fig:colliderplot} we plot the collider observables $ \delta_{Zh}$ and $\delta_{h^3}$ for different values of $N_s$ for the boundary of SNR and a set of parameter points with a SFOPT~\cite{Curtin:2016urg}  $v_c/T_c=0.6$.   As illustrated in Figure \ref{fig:colliderplot}, there can be a degeneracy in  $\delta \sigma_{Zh}$  even after the runs of the most optimistic future lepton collider proposals . If such a deviation were measured in $\delta \sigma_{Zh}$, it would likely be ascribed to a SFOPT while it could actually reflect no phase transition whatsoever.  It would then be necessary to then have a more precise triple Higgs coupling measurement at a future hadron collider to break the degeneracy between a SFOPT and SNR.  Alternatively, as previously mentioned one could also disentangle this possibility with the presence or absence of a stochastic gravitational wave signal at future gravitational wave observatories.

\begin{figure}[htbp] 
   \centering
   \includegraphics[width=3in]{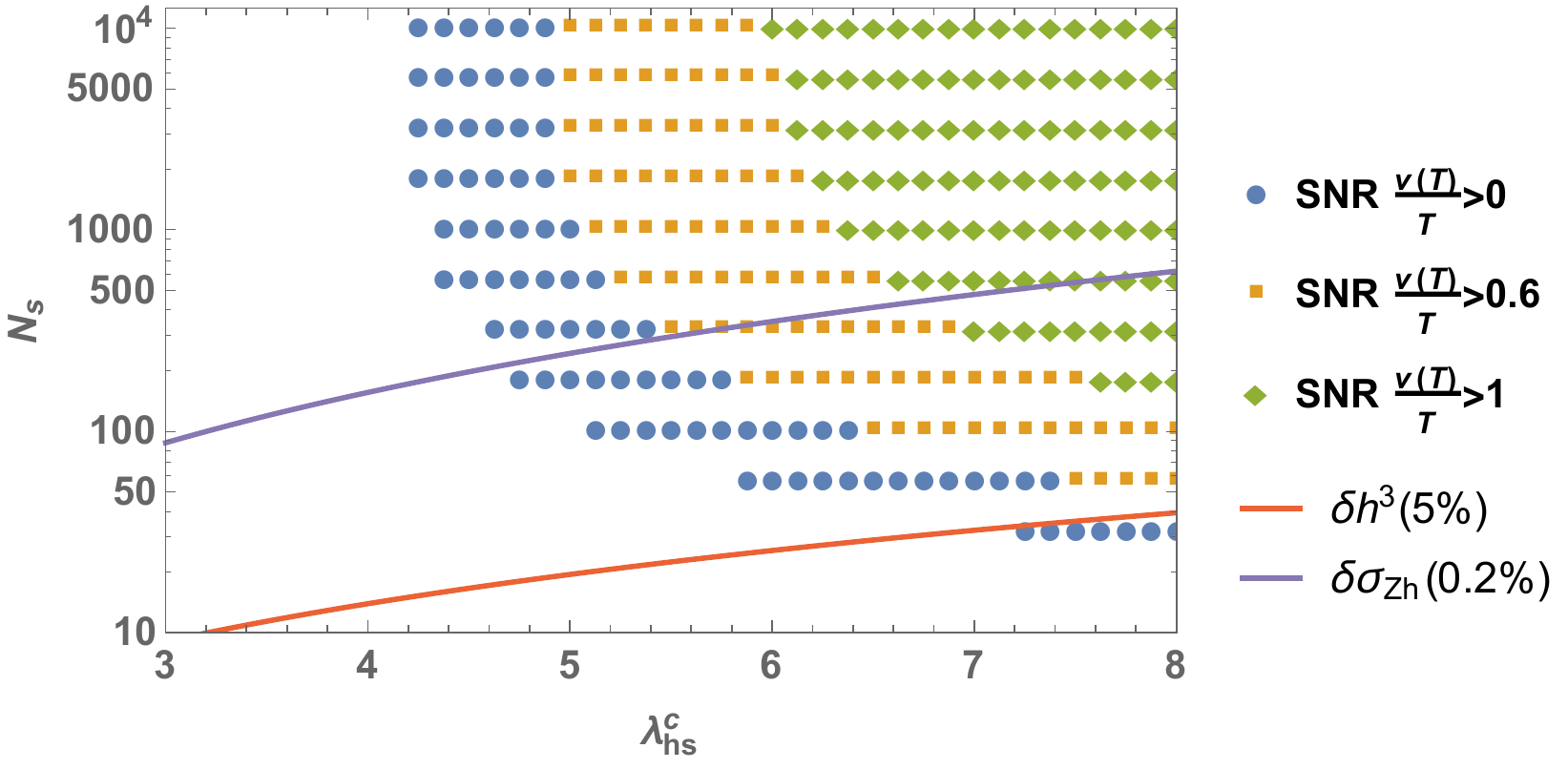} 
   \caption{Collider sensitivities from Figure \ref{fig:colliderplot} in the $N_s$ vs $\lambda_{hs}^c$  parameter space space for $m_s =100$ GeV superimposed with the SNR parameter space for $T=800$ GeV to avoid any finite-mass effects.}
   \label{fig:colliderplotB}
\end{figure}

For a given singlet mass, one can nonetheless set bounds on $N_s$ through precision Higgs measurements. In Figure \ref{fig:colliderplotB}, future collider bounds are shown for $m_s=100$ GeV that are overlaid with regions of SNR at high temperature.  Figure \ref{fig:colliderplotB} illustrates that precision measurement of $\delta \sigma _{Zh}$ to $0.2\%$ can set limits on $N_s$ to $\mathcal{O} (100)$. Furthermore, if one was interested in $\kappa$ being sufficiently high such that Sphalerons were too inefficient for baryogenesis this would provide even stronger constraints on allowed values of $N_s$(future colliders could constrain $\kappa >1$ models to $N_s > 500$).

\section{Conclusion}
We have outlined a scenario where EWSB is either persistent in the early universe or can go through a different order parameter history with respect to temperature.  There are a number of interesting potential cosmological consequences that we have outlined for baryogenesis and gravitational wave studies, but there are also inherently new cosmological phenomena as well.  For instance the novel equation of states which could lead to different structure formation or the alternative thermal decoupling and dark matter histories that are potentially present.  The cosmological effects in this model are generally small because they are coupled to EWSB which sets a scale in the problem.  However, just as SNR/ISB have been used for other applications~\cite{Mohapatra:1979qt,Mohapatra:1979vr,Dvali:1995cj,Salomonson:1984rh,Langacker:1980kd,Dvali:1995cc}, the cosmological effects could be much larger if ISB/SNR occurred in another sector where the scale and zero temperature couplings were not predetermined.  There are also additional collider signals that we have discussed and that can be correlated with a particular cosmology of interest.  

The phases we have shown in this letter for EWSB are robust in the large $N_s$ limit, but this is also inherently a regime where most collider and cosmological observables can vanish leaving an inherent ambiguity as to the history of our universe.  Therefore it is important to investigate other observables for this model, but also to further study this phase for smaller $N_s$ using non-perturbative lattice methods to understand the full parameter space.  It would also be useful to explore other non-perturbative methods for verification of SNR, such as those dealing with discrete anomaly matching~\cite{Komargodski:2017dmc}.  Regardless of the particular part of parameter space, the basic model outlined in this letter serves as an important alternative cosmological history that is consistent with all current and near future experimental searches. This model will hopefully serve as a benchmark for probing early universe cosmology as we plan for future experiments that look for verifiable information about our universe for $T\gtrsim$ a few GeV.

\section{Acknowledgements}
We would like to thank David Curtin, Daniel Egana-Ugrinovic, and Marilena LoVerde for useful discussions. The work of PM was supported in part by the National Science Foundation grant PHY-1620628.  PM would like to thank the Aspen Center for Physics, which is supported by National Science Foundation grant PHY-160761, where part of this work was completed.  HR is supported in part by the DOE under contract DEAC02-05CH11231.

\bibliographystyle{unsrt}
\bibliography{bibliography}{}

\end{document}